\documentclass{svproc}
\usepackage{url}

\usepackage{graphicx}
\usepackage{wrapfig}
\usepackage{lipsum}

\newcommand{\snn}{$\sqrt{s_{\mathrm{NN}}}$}

\newcommand{\dndy}{d$N$/d$y$}

\newcommand{\be}{$\begin{equation}$}
\newcommand{\ee}{$\end{equation}$}

\begin{document}

\mainmatter              
\title{Review of (anti-)Nuclei Production from High Energy Experiments}
\titlerunning{(Anti-)Nuclei Production}  
%
\author{Natasha Sharma}
\authorrunning{N. Sharma} 
%
%
\institute{Department of Physics, Panjab University, Chandigarh, India-160014. \\
\email{e-mail address: natasha.sharma@cern.ch} }

\maketitle              

\begin{abstract}
An overview of nuclei and anti-nuclei production with results from different experiments are discussed. The comparison of data with the thermal and coalescence models  is also discussed to understand their production mechanisms in high energy collisions. 
\vspace{-0.3cm}
.
\end{abstract}
\vspace{-0.7cm}
\section{Introduction}
\vspace{-0.3cm}
The Universe started with a big-bang with nearly equal abundance of matter and anti-matter. However, this symmetry got lost in the evolution of the universe with no visible amounts of anti-matter being present. 
The production of light (anti-)nuclei including (anti-)hypernuclei in high energy heavy-ion  collisions may be related to the matter-antimatter symmetry. 
The production of light (anti-)nuclei has been studied in the vast energy range in heavy-ion collisions experiments at Bevalac, AGS, SPS, RHIC, and LHC.
The two proposed production mechanisms for the (anti-)nuclei 
are the coalescence model
~\cite{Butler:1963pp} 
and the statistical thermal model
~\cite{Andronic:2010qu}. 	
The simple coalescence model assumes that \mbox{(anti-)}nuclei are formed if the constituents baryons are close in the coordinate and the momentum phase space.
The thermal model predicts the dependence of the particle yields of mass $m$ on the baryon chemical potential ($\mu_B$) and chemical freeze-out temperature ($T_{\rm chem}$) by the relation \dndy\ $\propto$ $\exp((\mu_B-m)/T_{\rm chem})$. 

\vspace{-0.4cm}

\section{Results and Discussions}
\begin{figure}
\vspace{-0.8cm}
\begin{center}
\includegraphics[width=5.1cm]{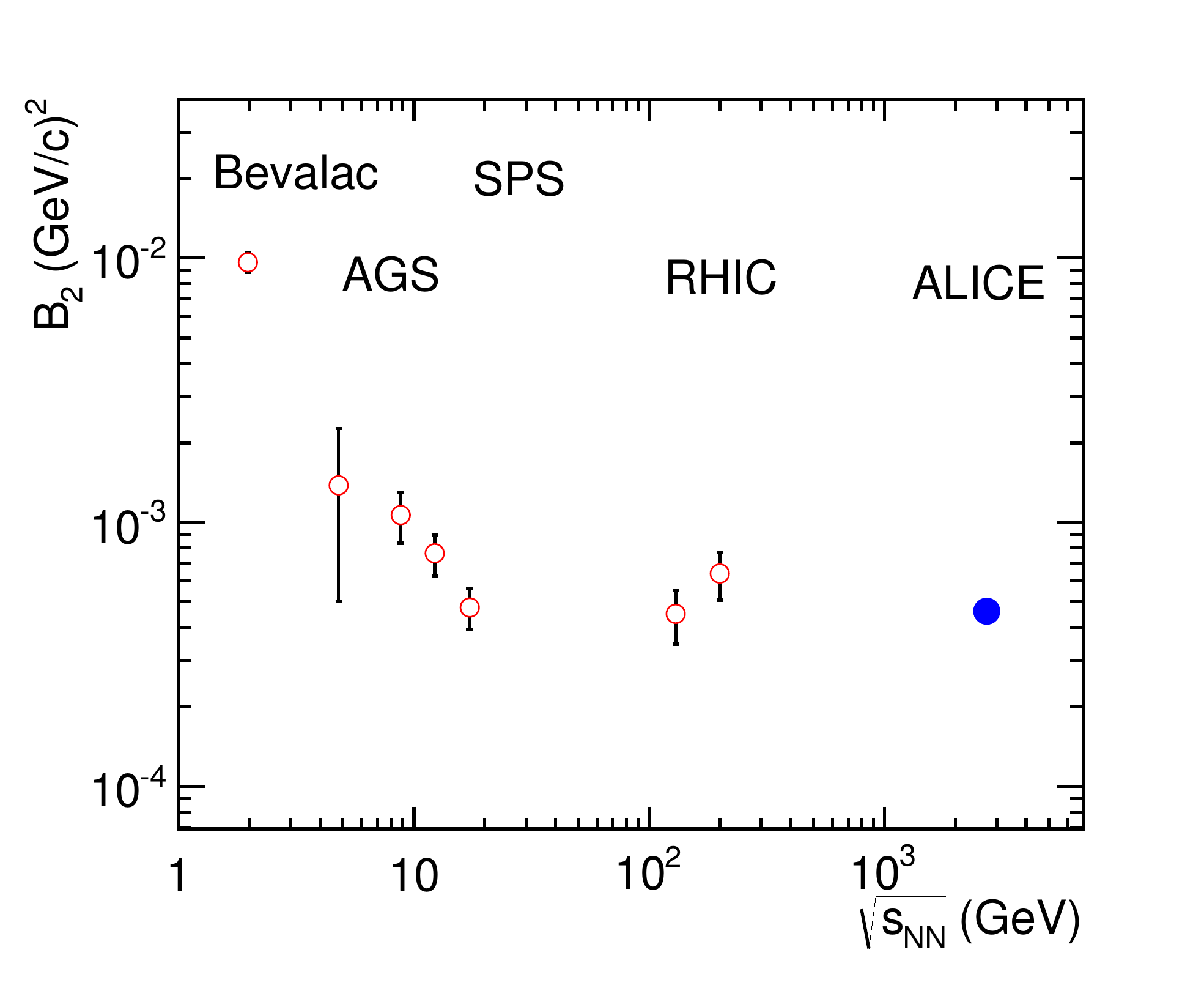}
\includegraphics[width=4.7cm]{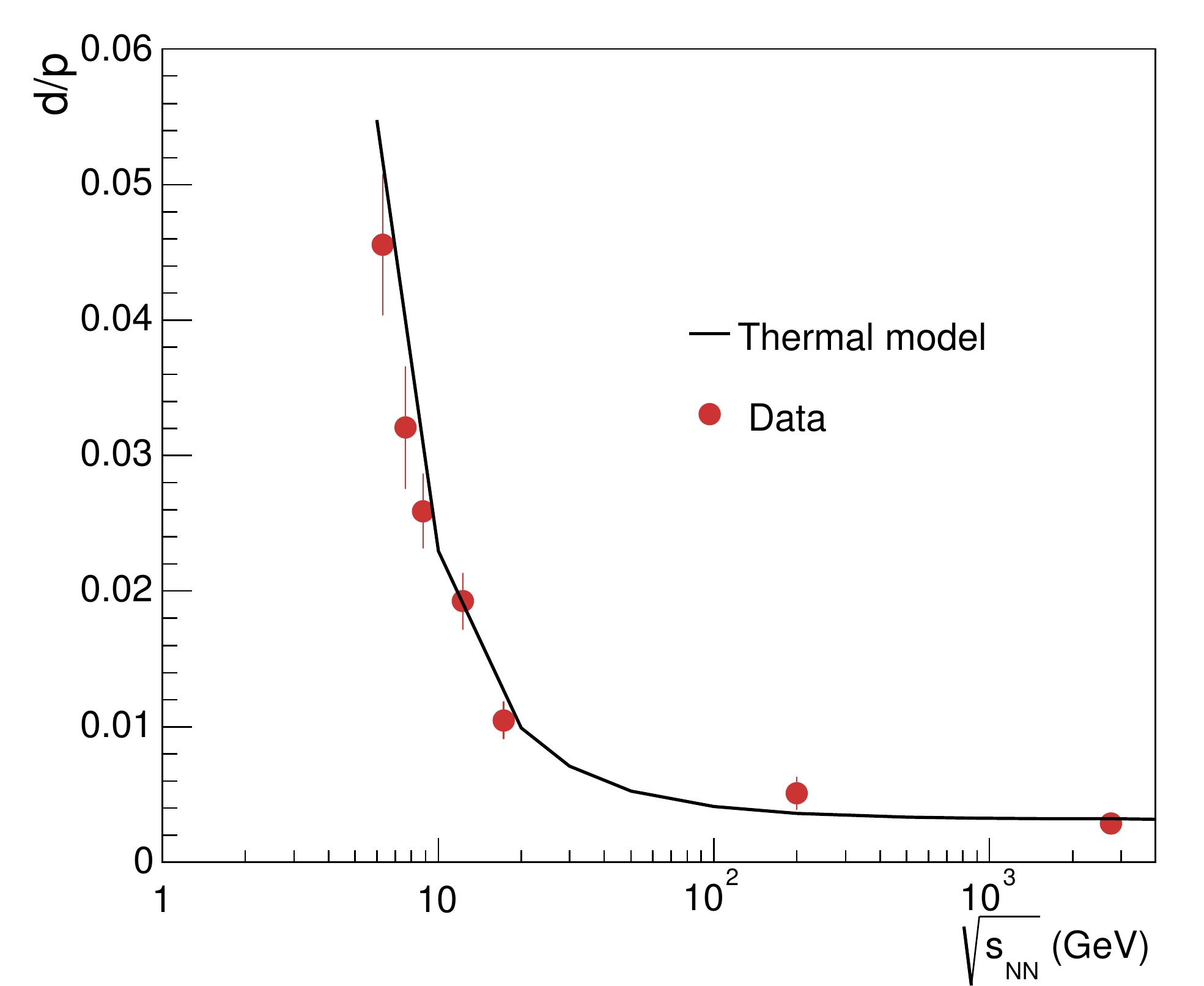}
\vspace{-0.3cm}
\caption{Left: B$_2$ measured from various experiments as a function of \snn\ ; 
Right: d/p ratio from data and thermal model prediction as a function of \snn.}
\label{fig1}
\end{center}
\vspace{-0.9cm}
\end{figure}

\vspace{-0.4cm}
In coalescence mechanism, the spectral distribution of the deuterons is related to that of primordial protons via
$E_d \, \frac{{\rm d^3} N_d}{{\rm d} p_d^3} =  B_2 \, \left(
E_{\rm p} \, \frac{{\rm d^3} N_{\rm p}}{{\rm d} p_{\rm p}^3}
\right)^2$,
assuming that protons and neutrons have the same momentum
distribution. $B_2$ is the coalescence parameter for deuteron with momentum of $p_d = 2 \, p_{\rm p}$.
Figure~\ref{fig1} (left) shows the energy dependence of the coalescence parameter $B_2$. It decreases from Bevalac to SPS energies, then remains almost constant up to RHIC energy~\cite{Adler:2004uy}. From RHIC top energy to LHC energy $B_2$ shows slight decrease~\cite{Adam:2015vda}. In the coalescence picture, this behavior is explained by an increase in the source volume i.e the larger the distance between protons and neutrons which are created in the collision, the less likely is that they coalesce. 
At RHIC energies, it is observed that elliptic flow ($v_2$) of light (anti-)nuclei when scaled with mass number follows $v_2$ of (anti-)protons suggesting that light (anti-)nuclei are produced via coalescence mechanism~\cite{Adamczyk:2016gfs}. However, the preliminary results from ALICE suggest that simple coalescence model fail to describe (anti-)deuteron $v_2$ at LHC energy~\cite{Lea:2016jus}.

The right plot of Fig.~\ref{fig1} shows the energy dependence of experimentally measured deuteron to proton ratio in heavy-ion collisions~\cite{Adam:2015vda,Anticic:2016ckv}. The comparison is made with the thermal model predictions which seems to explain the data well. It is observed that for a given collision energy,  $T_{\rm chem}$ and $\mu_B$ remain same for heavy-ion and small collisions system. At LHC, the d/p ratio for   p--Pb and pp collisions is smaller than in Pb--Pb system~\cite{Sharma:2016vpz}.  This observation could not be explained by the thermal model.

\begin{figure}
 \begin{center}
\includegraphics[width=4.7cm]{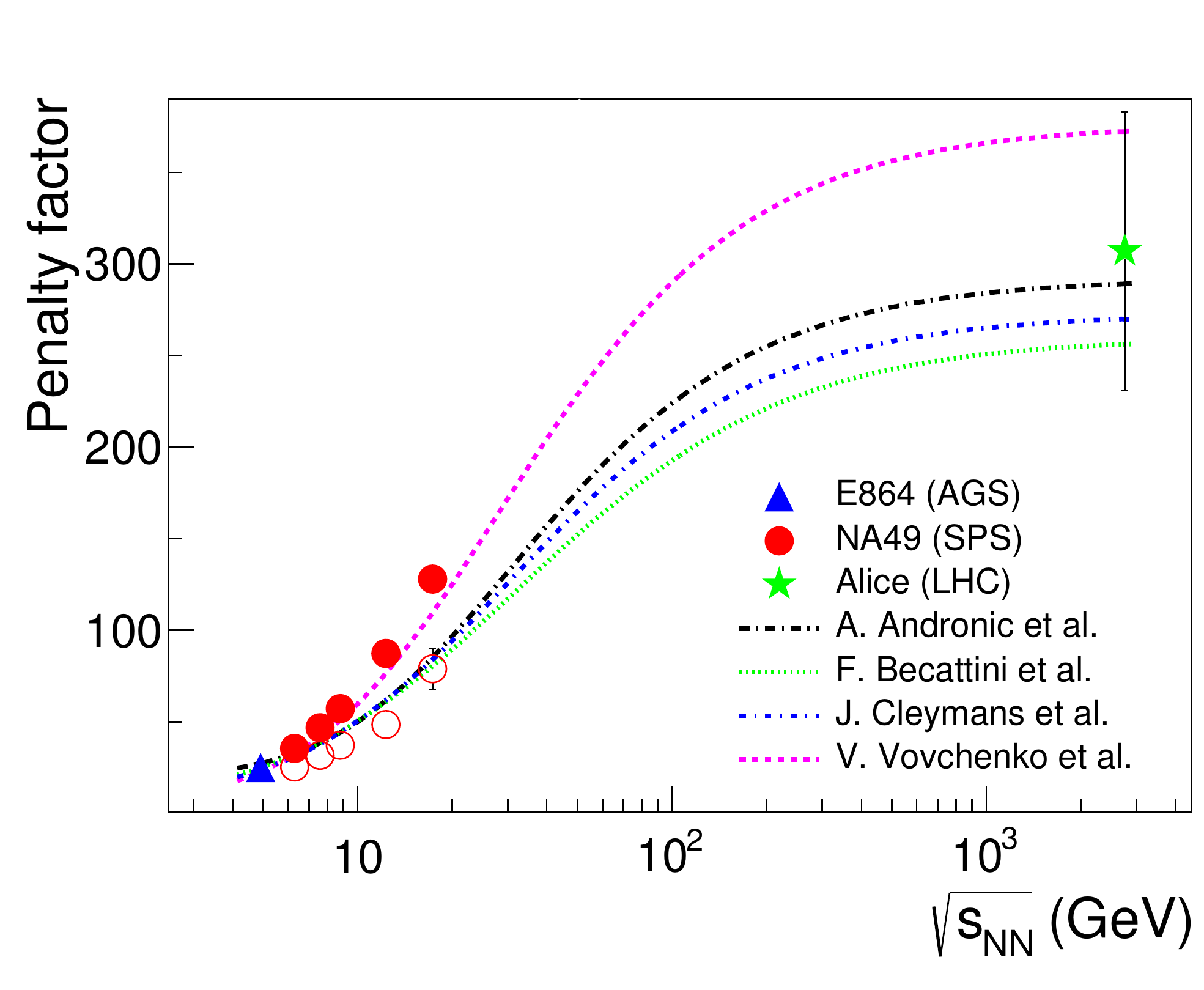}
\includegraphics[width=4.5cm]{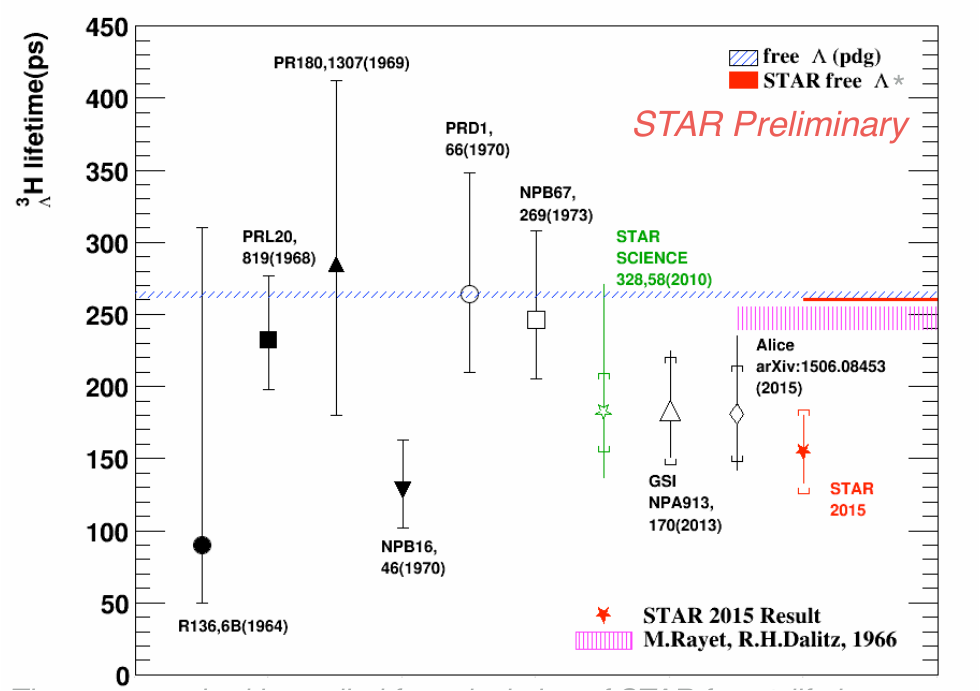}
\vspace{-0.3cm}
\caption{Left: Penalty factor as a function of energy for heavy-ion collisions.
Right: Lifetime measurements of hypertriton from different experiments.}
\label{fig2}
 \end{center}
\vspace{-0.8cm}
\end{figure}

For a given system and energy, the thermal model suggests that the nuclei yields follow an exponential decrease with the mass which is consistent with the experimental results. One has to pay a "penalty factor" to produce the heavier nuclei or each added nucleon~\cite{Sharma:2016vpz}. Figure~\ref{fig2} (left) shows the penalty factor as a function of energy for heavy-ion collisions~\cite{Anticic:2016ckv}. The markers show the experimental results (solid symbols for nuclei and open symbols for anti-nuclei) while the curves represent the thermal model predictions. The thermal model predictions are consistent with the heavy-ion experimental results within the uncertainities. However, this is not true for the smaller systems. For example, it is observed that penalty factor is about 600 for p--Pb collisions at LHC~\cite{Sharma:2016vpz} which is well above the thermal model expectation at this energy. 
This deviation from thermal model expectation in smaller collision systems needs theoretical explanation.

The right plot of Fig.~\ref{fig2} shows the experimental measurements of hyper-triton ($^{3}_{\Lambda}$H) lifetime from various experiments~\cite{Yifei:hyp2015}. The figure also shows the lifetime of the free $\Lambda$ from the PDG and measurement from STAR.  The recent experimental results from STAR~\cite{Yifei:hyp2015} and ALICE~\cite{Adam:2015yta} suggest that the $^{3}_{\Lambda}$H lifetime is less than that of free $\Lambda$. It will be interested to understand the theoretical explanation of the observed difference in lifetime of $^{3}_{\Lambda}$H and free $\Lambda$.

Recently, the ALICE experiment has measured the mass and binding energies of nuclei and anti-nuclei and found it to be compatible within uncertainties. This confirm the CPT invariance of light nuclei~\cite{Adam:2015pna}. 
The Quantum Chromodynamics (QCD) predicts the existence of exotic bound states of baryons~\cite{Adam:2015nca}. The thermal and coalescence models explain well the experimentally measured yields of light nuclei and hypertriton~\cite{Adam:2015vda}.  These models may predict the yields of exotic bound states and hence may provide baseline to test the existence of these states.
Various experiments attempt to search for these weakly decaying bound state of baryons. 
Recently, ALICE has tried to search for the $\Lambda \Lambda$ and $\overline{\rm \Lambda n}$ bound states~\cite{Adam:2015nca}. No evidence for these bound states is observed. Theoretical explanation is needed to understand the non-observation of these states.
\vspace{-0.5cm}

\section{Summary}
\vspace{-0.3cm}
Light (anti-)nuclei measurements have been performed by various experiments. The thermal model seems to describe (anti-)(hyper-) nuclei production well in heavy-ion collisions. However, deviation from thermal model is seen for small collision systems, the simple coalescence model could explain (anti-)nuclei production in small systems.
   More work is needed on theoretical part to understand the production mechanism of nuclei and anti-nuclei in high energy collisions.

\vspace{-0.3cm}
\paragraph{Acknowledgement} 
This work is supported by DST-SERB Ramanujan Fellowship grant no. SB/S2/RJN-084/2015. NS thanks Dr. L. Kumar for the discussions.
\vspace{-0.5cm}
%

%
%
%
%
%
%
%

%
%
\end{document}